\documentclass[twocolumn,showpacs,prb]{revtex4}

\usepackage{graphicx}

\begin{document}

\title{Magnetic Field Scaling in Spin Glasses \\
and the Mean-Field Theory}
\author{V.~S.~Zotev}
\author{R.~Orbach}
\affiliation{Department of Physics, University of California,
Riverside, California 92521}
\date{Submitted to PRB on January 14, 2002}

\begin{abstract}

The scaling of the magnetic field dependence of the remanent
magnetization for different temperatures and different spin-glass
samples is studied. Particular attention is paid to the effect of
the de Almeida-Thouless (AT) critical line on spin-glass dynamics.
It is shown that results of the mean-field theory of aging
phenomena, with two additional experimentally justified
assumptions, predict $H/H_{AT}(T)$ scaling for remanent
magnetization curves. Experiments on a single crystal Cu:Mn 1.5 at
\% sample in the temperature interval from $0.7T_{g}$ to
$0.85T_{g}$ give results consistent with this scaling.
Magnetization vs. field curves for different Cu:Mn and thiospinel
samples also scale together. These experimental results support
the predictions of the mean-field theory of aging phenomena.

\end{abstract}


\maketitle

\section{\label{intro}Introduction}

Effect of a magnetic field on the spin-glass state is one of the
most important open problems in spin-glass physics. Two major
theoretical descriptions of spin-glass phenomena have evolved over
the past twenty years. One of them is the mean-field theory for
the static and dynamical properties of spin glasses, based on the
Parisi replica-symmetry-breaking formalism and related
ideas.\cite{mez87,bou97} The alternative approach is the droplet
model, based on the Migdal-Kadanoff
approximation.\cite{fis86,mig75} The two pictures provide very
different physical interpretations of observable spin-glass
phenomena. This difference is particularly pronounced when
spin-glass properties in a magnetic field are considered. The
mean-field theory predicts a spin-glass state with
replica-symmetry breaking at finite magnetic fields below a
critical line in the $(T,H)$ plane.\cite{alm78} The droplet model
states that a true phase transition occurs at zero magnetic field
only. Compelling experimental support for either model has not yet
been presented. A detailed analysis of magnetic field effects on
real spin glasses can provide information about the comparative
validity of the theoretical predictions.

Recent experimental results favor the mean-field picture. Torque
measurements have shown that Heisenberg spin glasses with random
anisotropy are characterized by a true spin-glass ordered phase at
high magnetic fields.\cite{pet99} Experimental studies of
violations of the fluctuation-dissipation theorem under an
increasing field change also support predictions of the mean-field
theory.\cite{zot02}

The remanent magnetization, measured after a change in magnetic
field, contains all essential information about spin-glass
dynamics. The general features of its field dependence have been
studied for various spin glasses.\cite{tho74} All experimental
results to date, however, have been treated phenomenologically. A
comprehensive theoretical picture which can explain these results,
and predict the field dependence of measurable quantities over a
wide range of field variations, remains lacking. The mean-field
theory of aging phenomena,\cite{bou97,fra98,mar98,cug99} developed
in recent years, now appears able to provide such a description
within the linear response regime. The theory relates the
macroscopic relaxation properties of the spin-glass state to
microscopic correlations. Many conclusions, derived from this
theoretical picture, can be tested experimentally. In the present
paper, we study the scaling of magnetization curves with magnetic
field for several spin-glass samples. We show that, under some
additional experimentally justified assumptions, our experimental
results support predictions of this theory.

This paper is organized as follows. In the next Section, the
theoretical picture underlying our analysis is outlined. Sec.
III.A presents experimental results on the field scaling for
different temperatures. In Sec. III.B, experimental data for
different samples are compared. Section IV summarizes our
conclusions.

\section{\label{theor}Theoretical background and scaling predictions}

The \textit{equilibrium} susceptibility of an Ising spin glass,
identified with the equilibrium value of the experimental
field-cooled susceptibility, is given by the following expression:
\cite{par83}
\begin{equation}
\chi_{FC}=[1-\int_{0}^{1}q(x)dx]/T~~. \label{xfc}
\end{equation}
This susceptibility includes contributions from different pure
equilibrium states with the nontrivial distribution of overlaps
$q(x)$. The value $q(1)=q_{EA}$ is the equilibrium
Edwards-Anderson order parameter, and $q(0)=q_{min}$ is the
minimum possible overlap, nonzero in the presence of a magnetic
field. The spin-glass state is chaotic in magnetic field,
\cite{par83} meaning that the average equilibrium overlap of two
states at slightly different values of magnetic field is equal to
$q_{min}$.

The \textit{linear response} susceptibility, identified with the
experimental zero-field-cooled susceptibility at short observation
times $\tau=t-t_{w}$, is given by the fluctuation-dissipation
theorem (FDT) in its integral form: \cite{cug99}
\begin{equation}
\chi(t,t_{w})=[1-C(t,t_{w})]/T~~. \label{fdt}
\end{equation}
Here, $C(t,t_{w})$ is the autocorrelation function for a system of
$N$ Ising spins, defined as follows:
\begin{equation}
C(t,t_{w})=(1/N) \sum_{i=1}^{N} \langle S_{i}(t)S_{i}(t_{w})
\rangle~~. \label{cor}
\end{equation}
The linear response susceptibility is associated with transitions
within a single pure state. The difference between the values of
the field-cooled and zero-field-cooled susceptibilities is a
manifestation of replica-symmetry breaking.\cite{par83}

Spin-glass dynamics is limited to a single ergodic component,
because the energy barriers, separating the pure equilibrium
states, are divergent in the thermodynamic limit. The long-time
dynamics within one pure state can be viewed as a series of
transitions from a trap to a deeper trap.\cite{bou97} The barriers
surrounding these traps are high, but finite, and ``traps
encountered at long times tend to increasingly resemble the actual
states contributing to the equilibrium''.\cite{cug94} This
interpretation leads to a description of asymptotic spin-glass
dynamics, algebraically similar to the static
replica-symmetry-breaking formalism.

Dynamical definitions of $q_{EA}$ and $q_{min}$ are given in terms
of the correlation function: \cite{bou97,cug99}
\begin{equation}
q_{EA}=\lim_{\tau \rightarrow \infty} \lim_{t_{w} \rightarrow
\infty} C(t_{w}+\tau, t_{w})~~; \label{qea}
\end{equation}
\begin{equation}
q_{min}=\lim_{\tau \rightarrow \infty} C(t_{w}+\tau,t_{w})~~.
\label{qmi}
\end{equation}
Eq.~(5) means that, after a small field change following a finite
waiting time $t_{w}$, the system evolves towards an equilibrium
state that has the minimum possible correlation with the initial
state state at $t=t_{w}$.

Experiments \cite{oci85} and computer simulations \cite{and92}
suggest the following picture of spin-glass relaxation. At short
observation times, $\tau \ll t_{w}$, the relaxation is fast (on
the linear time scale) and equilibrium in nature. The correlation
function, Eq.~(3), drops from 1 to $q_{EA}$. The
fluctuation-dissipation theorem holds, and the susceptibility is
given by Eq.~(2). At longer observation times, $\tau > t_{w}$, the
relaxation is very slow and $t_{w}$-dependent. The correlation
decreases from $q_{EA}$ to $q_{min}$. The FDT is violated, and the
zero-field-cooled susceptibility relaxes towards the equilibrium
value, presumably given by Eq.~(1).

It is proposed in the mean-field theory of aging
phenomena\cite{bou97} that, for large $t_{w}$, the susceptibility
depends on its time arguments only through the correlation
function, i.e. $\chi=\chi[C(t,t_{w})]$, even when the
fluctuation-dissipation theorem is violated. The susceptibility
$\chi(C)$ is a piecewise function.\cite{mar98,cug99} It is linear
in the equilibrium regime:
\begin{equation}
\chi(C)=[1-C]/T \mbox{~,~~~} q_{EA}\leq C < 1~~. \label{sta}
\end{equation}
In the aging regime, the relaxing part of the susceptibility,
$\chi_{ag}(C)$, is nonlinear:
\begin{equation}
\chi(C)= [1-q_{EA}]/T+\chi_{ag}(C) \mbox{~,~~~} q_{min} < C <
q_{EA}~~. \label{age}
\end{equation}

The well-known Parisi-Toulouse approximation \cite{par80} makes
use of the following assumptions: the equilibrium susceptibility,
Eq.~(1), is independent of temperature, while $q_{EA}$ and
$q_{min}$ are functions of only temperature and magnetic field,
respectively. The dynamical version of this approximation implies
\cite{mar98,cug99} that the function $\chi_{ag}(C)$ in Eq.~(7) is
both $T$- and $H$-independent. This means that the dependence
$\chi(C)$ is universal in the aging regime, and follows a master
curve $\tilde{\chi}(C)$. If the value of the susceptibility at the
limit of validity of the FDT, i.e. at $C=q_{EA}$, is denoted as
$\chi_{ZFC}$, one can write the following:
\begin{equation}
\chi_{FC}=\chi_{FC}(H)~~;~~~~~~~~~~~\chi_{ZFC}=\chi_{ZFC}(T)~~.
\label{sim}
\end{equation}
Fig.~1, taken directly from Cugliandolo \textit{et al.},
\cite{cug99} displays the master curve $\tilde{\chi}(C)$. Each
point on this curve corresponds to a transition from the
equilibrium to the aging regime at some temperature $0<T<T_{g}$.
The quantity $q_{d}$ is the initial correlation $C(t_{w},t_{w})$,
which depends on the number of spin components. It appears instead
of unity in Eqs.~(6) and (7) if the spins are not Ising.

\begin{figure}
\resizebox{\columnwidth}{!}{\includegraphics{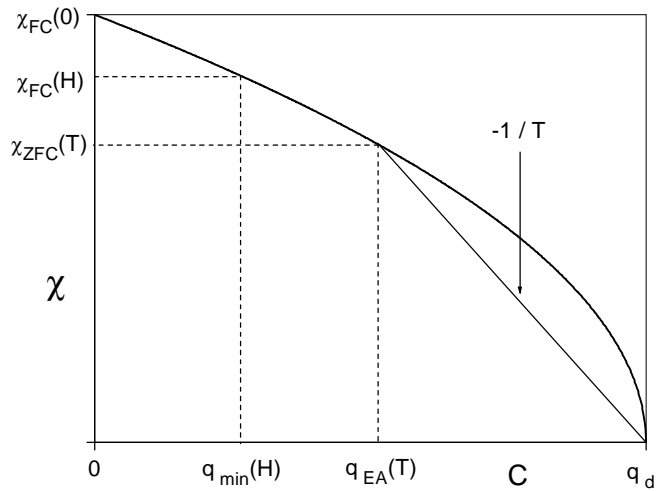}}
\caption{\label{Fig1} A diagram of the spin-glass relaxation at
temperature $T$ after magnetic field $H$ is applied. The thick
line is the master curve $\tilde{\chi}(C)$. The straight line
segment from $(q_{d},0)$ to $(q_{EA},\chi_{ZFC})$ represents the
equilibrium relaxation regime. The slope is $-1/T$. The master
curve segment from $(q_{EA},\chi_{ZFC})$ to $(q_{min},\chi_{FC})$
corresponds to the aging regime. }
\end{figure}

The contribution of this paper is the experimental study of the
magnetic field dependence of the remanent susceptibility
$\chi_{FC}(H)-\chi_{ZFC}(T)$. According to Fig.~1, it is related
to the difference $q_{EA}(T)-q_{min}(H)$. In order to derive a
magnetic field scaling relationship, we must introduce two
additional assumptions. First, we consider relatively high
temperatures and assume that the master curve $\tilde{\chi}(C)$ at
low $C$ can be approximated by a straight line. Then the triangles
in Fig.~1 are geometrically similar for all allowed $T$ and $H$,
and the following relation holds:
\begin{equation}
\frac{\chi_{FC}(H)-\chi_{ZFC}(T)}{\chi_{FC}(0)-\chi_{ZFC}(T)}=
\frac{q_{EA}(T)-q_{min}(H)}{q_{EA}(T)}~~. \label{rat}
\end{equation}
Second, let us suppose that $q_{min}(H)$ is a homogeneous function
of order $p$, that is $q_{min}(aH)=a^{p}q_{min}(H)$ with some $p
\neq 0$. Then, introducing the critical AT field, one can write:
\begin{equation}
\frac{q_{min}(H)}{q_{EA}(T)}=\frac{q_{min}(H)}{q_{min}[H_{AT}(T)]}=
\frac{q_{min}[H/H_{AT}(T)]}{q_{min}(1)}~~. \label{sca}
\end{equation}
Here we used the condition\cite{par83} that the de
Almeida-Thouless critical line is defined by $q_{min}=q_{EA}$. It
follows from Eqs.~(9) and (10) that the remanent susceptibility,
$\chi_{FC}(H)-\chi_{ZFC}(T)$, should scale as $H/H_{AT}$. This is
a consequence of the proposed universality of $\tilde{\chi}(C)$.
The present paper is devoted to the experimental study of this
field scaling. We shall also use our experimental data to justify
the two assumptions which lead to Eqs.~(9) and (10).

\section{\label{exper}Experimental results and analysis}

Before presenting our experimental results, we would like to make
some preliminary remarks.

Predictions of the mean-field theory of aging phenomena, mentioned
in the previous section, are expected to hold only in the linear
response regime. This means that a change in magnetic field,
acting as a probe of the spin-glass state, must be much smaller
than the AT field at a given measurement temperature. A larger
field change would lead to a deviation from linear response. The
measured zero-field-cooled susceptibility would then become
field-dependent, and the arguments, based on Fig.~1, could not be
used.

Eqs.(8)-(10) can be applied to experimental data only if the
zero-field-cooled susceptibility is measured at the end of the
fluctuation-dissipation regime. For relatively short waiting
times, the transition from one regime to the other is not well
defined. Computer simulations show \cite{and92} that violation of
the FDT becomes visible at observation times $\tau$ at least one
order of magnitude shorter than the waiting time $t_{w}$, the
violation becoming strong at $\tau \approx t_{w}$. All of our
experiments have been performed on a commercial Quantum Design
SQUID magnetometer. The shortest possible observation time is
about $40~s$. The typical effective cooling time is $600~s$
because of the rather low cooling rate near the measurement
temperature. By the time the first experimental point is taken,
the fast initial decay is essentially over. Thus, the short-time
measurements yield results, which are approximately at the end of
the fluctuation-dissipation regime, even at zero waiting time.

In our analysis, we use experimental values of magnetizations
instead of susceptibilities. This is because numerical
differentiation requires fitting, and any fitting involves
interpretation. Of course, all arguments regarding the field
scaling apply to magnetizations as well. For example, Eqs.~(9) and
(10) suggest that the slope of the remanent magnetization,
$MFC-ZFC$, at field $H$, divided by its slope at $H=0$, will be a
function of $H/H_{AT}(T)$. The rest of this Section presents our
experimental results, which will support this prediction.

\subsection{\label{expA} Field scaling for different temperatures}

In order to test validity of Eqs.~(9) and (10), we measured the
field-cooled ($MFC$) and zero-field-cooled ($ZFC$) magnetizations
as functions of the field $H=\Delta H$ for four different
temperatures. All results reported in this subsection were
obtained for a single crystal Cu:Mn 1.5 at \%. It is a long-range
Heisenberg spin glass with a glass temperature $T_{g}$ of
approximately $15.2~K$. The experimental phase diagram for this
sample will be presented elsewhere. The measurements of the $ZFC$
magnetization were made at the shortest observation time, with
zero waiting time between the cooling and application of the
field.

\begin{figure}
\resizebox{\columnwidth}{!}{\includegraphics{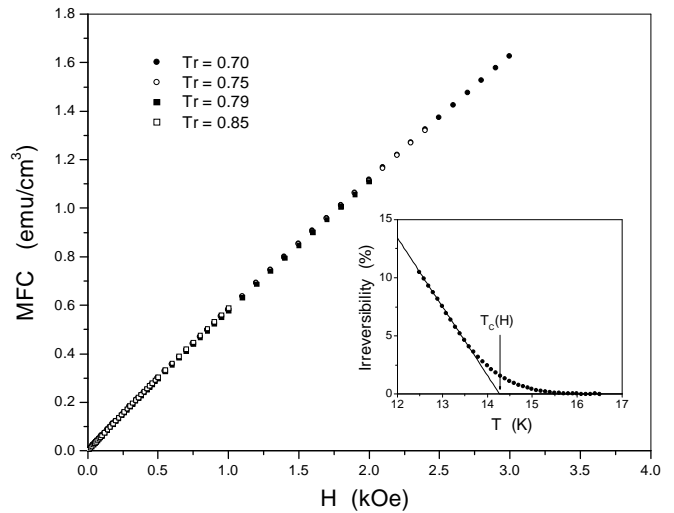}}
\caption{\label{Fig2} The field-cooled magnetization of the single
crystal Cu:Mn 1.5 at \% for different temperatures $Tr=T/Tg$. The
inset displays the irreversibility $(1-ZFC/MFC)*100\%$ as a
function of temperature for $H=200~Oe$.}
\end{figure}

Fig.~2 exhibits the field dependence of the field-cooled
magnetization. The data are presented as taken, without any
rescaling of the field or adjustment of the magnetization
magnitude. Error bars are considerably smaller than the symbol
sizes and are not shown in the figure. The same applies to the
other figures in this paper that do not exhibit error bars. It is
evident that $MFC(H)$ is virtually independent of temperature.
Therefore, the Parisi-Toulouse approximation works rather well in
the case of this sample. This appears to be a common feature of
Cu:Mn spin glasses.\cite{cug99}

The spin-glass phase is characterized by the difference between
the field-cooled and zero-field-cooled magnetizations. The inset
of Fig.~2 displays temperature dependence of the $MFC-ZFC$
irreversibility at fixed magnetic field. Weak irreversibility in
the Cu:Mn 1.5 at \% single crystal appears slightly above the
glass temperature of $15.2~K$. The irreversibility becomes strong
as temperature is lowered, evolving to a region where it increases
linearly with a large slope. The strong $MFC-ZFC$ irreversibility
is interpreted as a sign of the spin-glass phase transition. The
$T$-intercept of the linear fit in this region is taken as the
crossover temperature $T_{c}(H)$. These crossover temperatures,
determined for different magnetic fields $H$, define the AT line.

Fig.~3 displays field dependences of the $MFC-ZFC$ irreversibility
for different temperatures. The four curves are similar in shape,
but the field scale for each depends on the measurement
temperature. The peak in $MFC-ZFC$ corresponds to a strong
violation of linear response. It turns out that the position of
this peak depends on the value of the AT field.

\begin{figure}
\resizebox{\columnwidth}{!}{\includegraphics{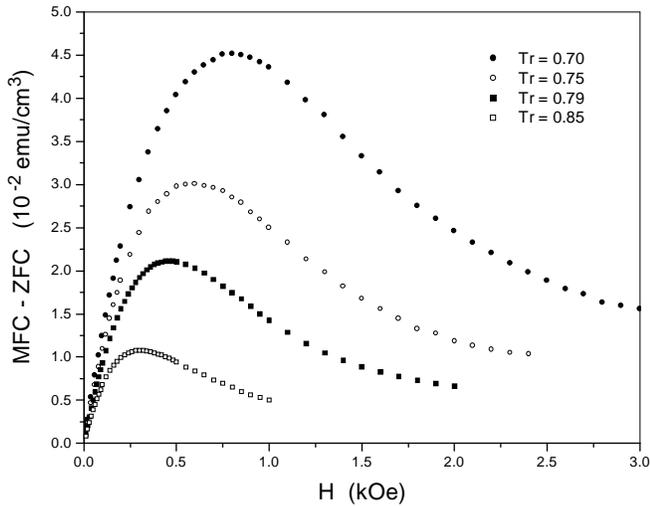}}
\caption{\label{Fig3} The remanent magnetization of the single
crystal Cu:Mn 1.5 at \% as a function of field for the same
temperatures as in Fig.~2.}
\end{figure}

\begin{figure}
\resizebox{\columnwidth}{!}{\includegraphics{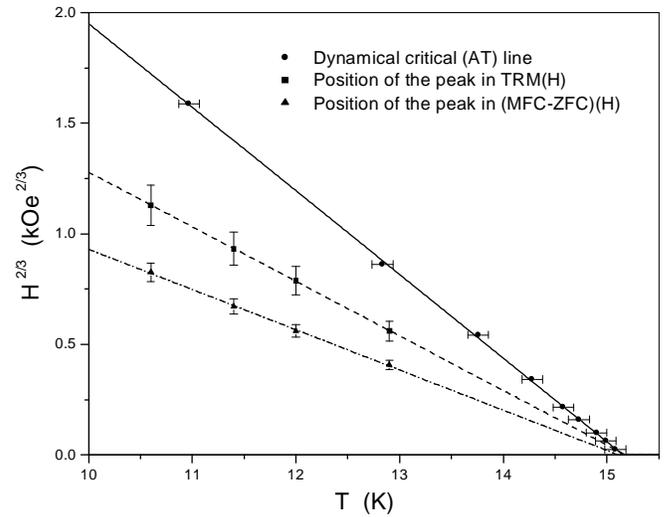}}
\caption{\label{Fig4} From top to bottom: $H_{AT}(T)$, the
experimental AT critical line; $H_{M}(T)$, the position of the
peak in $TRM(H)$; $H_{m}(T)$, the position of the peak in
$(MFC-ZFC)(H)$.}
\end{figure}

\begin{figure}
\resizebox{\columnwidth}{!}{\includegraphics{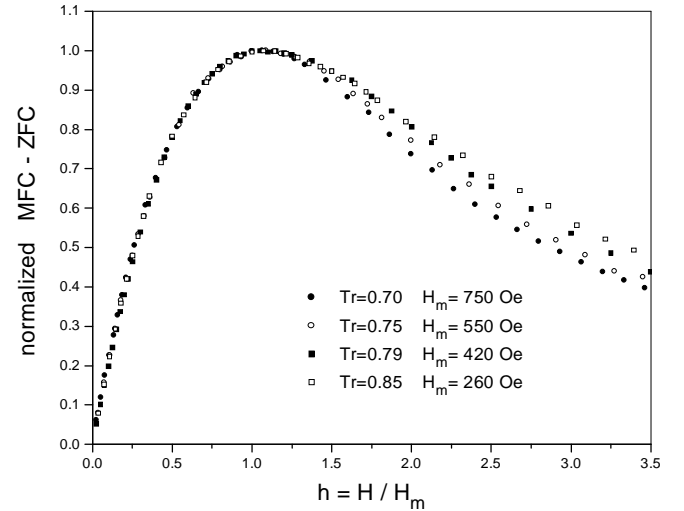}}
\caption{\label{Fig5} Scaling of the remanent magnetization curves
for different temperatures. The positions of the peaks,
$H_{m}(T)$, are used as scaling parameters.}
\end{figure}

The experimental AT line for the Cu:Mn 1.5 at \% single crystal
sample is exhibited in Fig.~4. One can see that this line has the
functional form $T_{g}-T_{c}(H) \propto H^{2/3}$, typical of the
AT line \cite{alm78} in the Sherrington-Kirkpatrick (SK) model.
\cite{she75} Fig.~4 also shows temperature dependence of the
position of the peak in $TRM(H)$, which we denote as $H_{M}(T)$,
and of the peak in $(MFC-ZFC)(H)$, which we refer to as
$H_{m}(T)$. The thermoremanent magnetization, $TRM$, is not equal
to $MFC-ZFC$ if linear response is violated, and $H_{m} < H_{M}$
for any temperature. However, both $H_{M}(T)$ and $H_{m}(T)$ are
proportional to the critical field $H_{AT}(T)$, according to
Fig.~4. The peak in $(MFC-ZFC)(H)$ is easier to identify than the
AT line itself. In this subsection, we shall use its position,
$H_{m}(T)$, as the field scaling parameter, and consider $H/H_{m}$
scaling instead of $H/H_{AT}$ scaling.

Fig.~5 exhibits the $MFC-ZFC$ irreversibility, normalized by its
value at $H_{m}$, and plotted versus $h=H/H_{m}$. One can see that
the curves for different temperatures fall on top of one another
for $h<1$. This is the interval of field variations where linear
response holds, at least approximately. Therefore, the field
scaling, predicted by Eqs.~(9) and (10), is indeed observed in
spin-glass experiments. For $h>1$, however, the scaling is rather
poor. There is strong nonlinearity in the spin-glass response in
this regime, and the mean-field theory of aging phenomena will not
be applicable.

Of course, nonlinearity by itself does not necessitate lack of
scaling. There are at least two probable reasons for the observed
differences in high-field behavior at different temperatures.
First, freezing of the transverse spin components \cite{gab81} in
Heisenberg spin glasses produces the weak irreversibility in the
longitudinal direction. Second, the distribution of glass
temperatures, always present in real samples, has a stronger
effect at higher measurement temperatures and higher fields. The
inset of Fig~2 shows that there is a significant remanence at the
crossover temperature. It corresponds to the remanence near the AT
line at $h \approx 3.5$ in Fig.~5. Obviously, we cannot expect the
$H/H_{AT}$ scaling to hold precisely in this region.

The scaling of the $MFC-ZFC$ curves in Fig.~5 suggests, \textit{a
posteriori}, that the assumptions leading to Eqs.~(9) and (10) are
in fact reasonable. In order to clarify this conclusion, we
determined the master curve $\tilde{\chi}(C)$ according to the
method of Cugliandolo \textit{et al}.\cite{cug99} If the
zero-field cooled susceptibility is measured at the limit of
validity of the FDT, the corresponding correlation can be obtained
from Eq.~(6). Then the values of $\chi_{ZFC}(T)$ for different
temperatures, plotted vs. $C(T)$, span the master curve
$\tilde{\chi}(C)$. The experimental master curve is shown in
Fig.~6. The data points were taken in the interval from $T=2.4~K$
to $T=15.0~K$ at the same low field $H=16~Oe$. Each measurement
was independent of the others, and included a quench from above
the glass temperature.

Fig.~6 demonstrates that the experimental dependence of $\chi$ on
$C$ is close to linear over a wide range of correlations. This
justifies the assumption underlying Eq.~(9): the master curve
$\tilde{\chi}(C)$ at relatively low $C$ can be approximated by a
straight line.

\begin{figure}
\resizebox{\columnwidth}{!}{\includegraphics{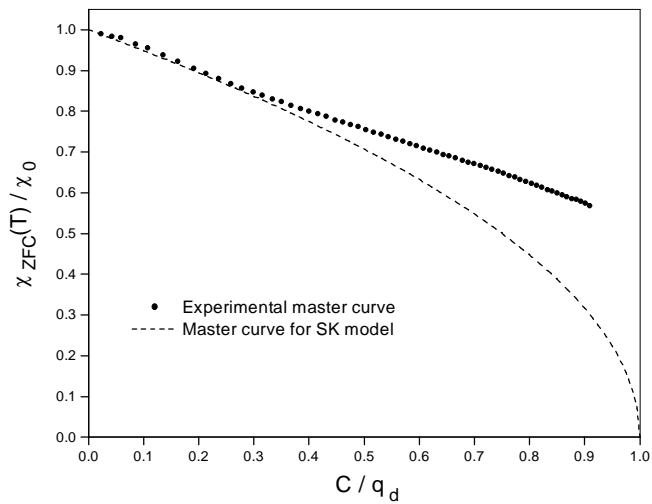}}
\caption{\label{Fig6} The master curve $\tilde{\chi}(C)$ for the
single crystal Cu:Mn 1.5 \%, estimated from the experimental
$\chi_{ZFC}(T)$ dependence. The master curve for the SK model is
$\tilde{\chi}=(1-C)^{1/2}$.}
\end{figure}

The experimental data for the AT critical line in Fig.~4 suggest
that, to the leading order of magnitude, $q_{min}(H) \propto
H^{2/3}$. Therefore, $q_{min}(H)$ is indeed a homogeneous function
of order $p=2/3$, and the assumption underlying Eq.~(10) is also
verified.

These arguments, of course, should be taken with caution. A
reliable experimental determination of the master curve
$\tilde{\chi}(C)$ would require independent measurements of both
the susceptibility and the correlation. Fig.~6 only gives an idea
of how the real curve may look. In particular, the experimental
master curve in Fig.~6 does not exhibit a significant downturn at
high values of $C/q_{d}$. The difference between this curve and
the theoretical curve $\tilde{\chi}=(1-C)^{1/2}$ for the SK model
\cite{mar98} is very pronounced in that region. This problem,
however, is beyond the scope of the present paper. It is also
evident that we cannot expect the above two assumptions to hold
beyond the leading order of magnitude even at high enough
temperatures and low values of $C$. It would be correct to say,
therefore, that the AT field defines a characteristic field scale
at any temperature, but the scaling itself is always approximate.

\subsection{\label{expB} Field scaling for different samples}

The results of the previous subsection suggest that the master
curve $\tilde{\chi}(C)$ in Fig.~1 is universal, that is $T$ and
$H$ independent, thus supporting the Parisi-Toulouse
approximation. It would be interesting to see, therefore, if this
curve depends on the choice of sample. Different spin-glass
samples have different microscopic properties, and, consequently,
different effective magnetic field scales. If the Parisi-Toulouse
approximation holds, the magnetic field $H$ appears in the
analysis through $q_{min}(H)$ only. Therefore, if the master curve
$\tilde{\chi}(C)$ is sample independent, and $q_{min}(H)$ has
always the same functional form, we can expect the scaling of
magnetization curves to hold for different samples.

In order to study this issue, we measured the field-cooled (MFC)
and the thermoremanent (TRM) magnetizations as functions of
$H=\Delta H$ for five different spin-glass samples. In addition to
the single crystal Cu:Mn 1.5 at \%, described previously, we used
a single crystal Cu:Mn 0.6 at \%, with the glass temperature of
about $6.0~K$. Both samples have been prepared in Kammerlingh
Onnes Laboratory (Leiden). Similar single crystals have been used
for newtron-scattering experiments. \cite{lam95} The other three
of our samples are polycrystalline. The polycrystal Cu:Mn 6.0 at
\% has been extensively studied before. \cite{ken91} Its glass
temperature is near $31.0~K$. The thiospinel
$CdCr_{1.7}In_{0.3}S_{4}$ is an insulating short-range spin glass
with $T_{g}=16.7~K$. It has also been studied in detail.
\cite{vin87} The second thiospinel sample, used in our analysis,
has been obtained from the part of the first sample by sifting it
through a $100~nm$ mesh. This was done to probe the finite-size
effects. \cite{swa00} The sifted thiospinel has a slightly lower
glass temperature $T_{g} \approx 16.5~K$.

Fig.~7 displays the thermoremanent magnetizations versus the field
$H$ for these five samples. All data points are taken at the same
short observation time of $90~s$ after the field is cut to zero.
The waiting time between the end of the cooling process and the
field change is $30~min$. The reduced measurement temperatures,
$Tr=T/T_{g}$, for different samples are not exactly the same, but
it is not very important considering the results of Sec.~III.A. It
is evident from Fig.~7 that different samples have very different
characteristic field scales. The effect of the same magnetic field
is strongest in the case of the thiospinel, and it is much less
pronounced for the single crystal Cu:Mn 1.5 at \%. The
characteristic field scales for these two samples differ by one
order of magnitude.

\begin{figure}
\resizebox{\columnwidth}{!}{\includegraphics{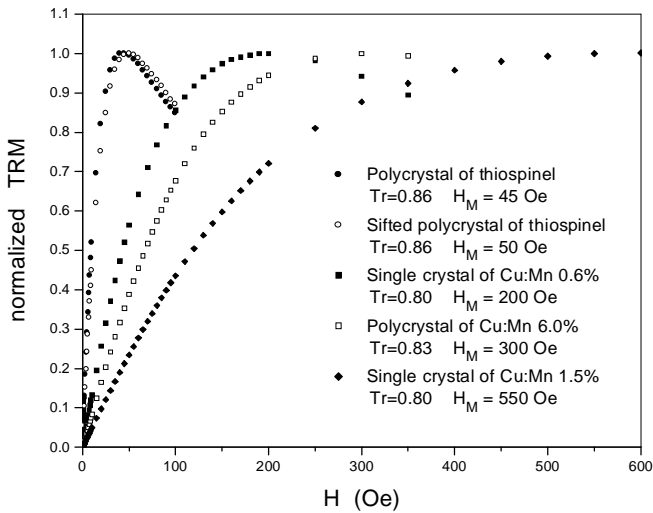}}
\caption{\label{Fig7} Thermoremanent magnetization for five
spin-glass samples measured at $t_{w}=30~min$. The positions of
the maxima, $H_{M}(T)$, are proportional to the AT fields.}
\end{figure}

\begin{figure}
\resizebox{\columnwidth}{!}{\includegraphics{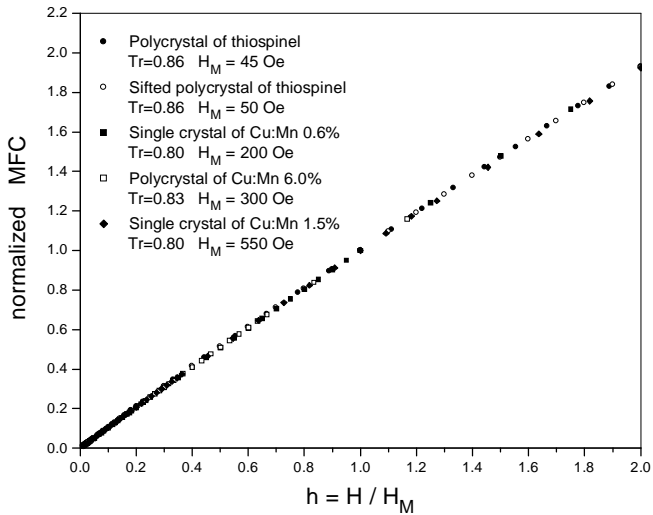}}
\caption{\label{Fig8} Scaling of the field-cooled magnetization
curves for the five samples. The values of $H_{M}(T)$ are used as
scaling parameters.}
\end{figure}

\begin{figure}
\resizebox{\columnwidth}{!}{\includegraphics{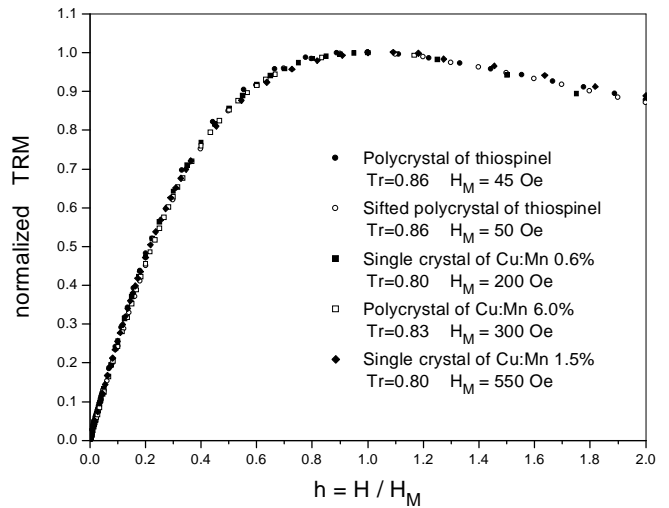}}
\caption{\label{Fig9} Scaling of the thermoremanent magnetization
curves from Fig.~7, plotted vs. $H/H_{M}$.}
\end{figure}

One can see from Fig.~7 that each $TRM(H)$ curve has a maximum.
According to Fig.~4, the position of this maximum, $H_{M}(T)$, is
proportional to the corresponding AT field, $H_{AT}(T)$. In this
subsection, we will characterize each sample by its value of
$H_{M}$, and consider $H/H_{M}$ scaling instead of the $H/H_{AT}$
scaling.

Fig.~8 exhibits experimental $MFC(H)$ curves, plotted versus
$h=H/H_{M}$ and normalized by their values at $h=1$. All five
curves seem to scale well together. This suggests that, apart from
the differences in the field scales, the physical mechanism behind
the observed field dependence is the same for all these very
different samples. Fig.~8 suggests that the functional form of
$q_{min}(H)$ is essentially independent of sample.

The experimental $TRM(H)$ curves from Fig.~7, plotted versus the
reduced field $h=H/H_{M}$, are presented in Fig.~9. The overall
scaling is surprisingly good, taking into account the diversity in
properties of the five samples. We conclude that the observed
nonlinearity in spin-glass response has the same physical origin
for all samples.

The major deviations from perfect scaling seem to result from
finite-size effects. They are not clearly visible in Fig.~9, but
can be seen in Fig.~10. This figure exhibits the normalized $TRM$
curves, divided by $h=H/H_{M}$, for the two thiospinel samples.
The full thiospinel sample with $T_{g} \approx 16.7~K$ has a broad
distribution of particle sizes. The sifted thiospinel sample with
$T_{g} \approx 16.5~K$ consists of particles smaller than $100~nm$
in diameter. The measurement temperatures were $14.4~K$ and
$14.2~K$, respectively, so that the reduced temperature
$Tr=T/T_{g}$ has the same value of 0.86 for both samples.

Two conclusions can be derived from the data in Fig.~10. First,
the characteristic field for the sifted sample is about 10\%
higher than for the full sample. This is consistent with the
mean-field-theory predictions. The differences in free energies
per site for different equilibrium states are of the order of
$1/N$, where $N$ is the total number of spins. For smaller $N$, a
stronger perturbation is needed to redistribute these energy
differences and thus reshuffle the weights of different pure
states.\cite{rit94} Therefore, nonlinearity in response appears at
higher field changes for smaller particles. The second conclusion
is that the differences in behavior between the full sample and
the sifted sample are more pronounced in the low-field region.
This is also in agreement with results of the mean-field theory.
The magnetic correlation length, $\xi_{H}$, drops sharply as the
change in field increases.\cite{rit94} If the field change makes
two spins uncorrelated, existence of a grain wall between them
becomes irrelevant. Thus, finite-size effects on spin-glass
dynamics are more pronounced at lower field changes. Of course,
there may be other reasons for the differences between the two
thiospinel samples.

\begin{figure}
\resizebox{\columnwidth}{!}{\includegraphics{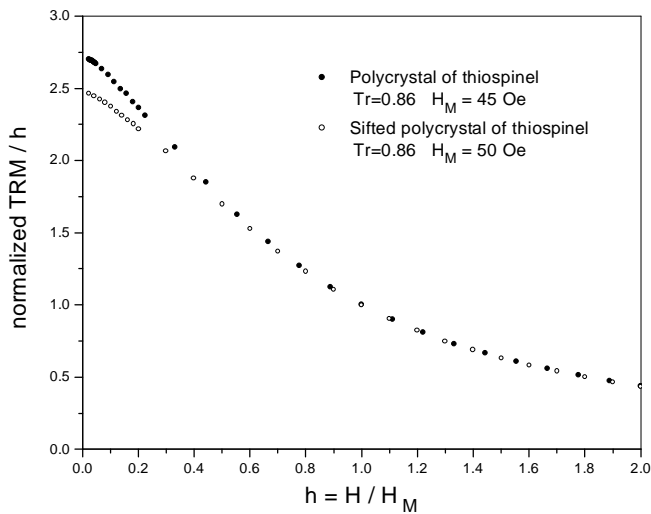}}
\caption{\label{Fig10} Comparison of results for the two
thiospinel samples. The differences in the values of $H_{M}$ and
in the low-field behavior can be attributed to the finite-size
effects on the spin-glass properties of the sifted sample.}
\end{figure}

Apart from the finite-size effects, the overall scaling in Fig.~9
is impressive, and demonstrates the validity of the theoretical
analysis in Section II. Our results suggest that both the master
curve $\tilde{\chi}(C)$ and the function $q_{min}(H/H_{AT})$ are
essentially independent of the nature of a particular spin-glass
sample. This is a strong evidence in support of mean-field theory.

\section{\label{concl}Conclusion}

The mean-field theory of aging phenomena establishes a relation
between the susceptibility $\chi$ and the correlation $C$ in both
the equilibrium and aging regimes of spin-glass relaxation. It
suggests that $\chi(C)$ is a universal function in the aging
regime, and that the remanent susceptibility
$\chi_{FC}-\chi_{ZFC}$ is related to the difference
$q_{EA}-q_{min}$ in the values of the correlation. We have shown
in this paper that two additional assumptions, the linearity of
$\chi(C)$ in the aging regime and the homogeneity of $q_{min}(H)$
in a wide range of fields, lead to a prediction of $H/H_{AT}(T)$
scaling of the remanent susceptibility curves. Our experiments on
the single crystal Cu:Mn 1.5 at \% demonstrate the existence of
such scaling in the temperature interval $T/Tg=0.7...0.85$.
Moreover, the thermoremanent magnetization curves for different
spin-glass samples also scale quite well together if plotted vs.
$H/H_{AT}$. These results indicate that there is universality for
the magnetic properties of different spin glass systems at
different temperatures. They also suggest that the magnetic field
effects on spin-glass dynamics in the linear response regime are
correctly described by the mean-field theory of aging phenomena.\\

We appreciate the courtesy of Professor~J.~A.~Mydosh from
Kammerlingh Onnes Laboratory (The Netherlands) who kindly provided
us with the Cu:Mn single crystal samples. We are also eternally
grateful to Dr.~J.~Hammann and Dr.~E.~Vincent from CEA Saclay
(France) for giving us one half of their thiospinel sample. We
thank Dr.~G.~G.~Kenning for his constant interest and support.

\end{document}